\documentclass{mem}
\usepackage{natbib}\usepackage{txfonts}\usepackage{balance}
\usepackage{graphicx}
\usepackage{txfonts}
\usepackage[a4paper]{hyperref}
\idline{79}{3}
\begin{document}
\def\teff{$T\rm_{eff }$}
\def\kms{$\mathrm {km s}^{-1}$}

\title{
Multimodal Horizontal Branches: Empirical Evidence and Possible Evolutionary Scenarios
}


\author{
M. Catelan 
          }


\institute{
Pontificia Universidad Cat\'olica de Chile, Departamento de Astronom\'{i}a y 
Astrof\'{i}sica,\\ Av. Vicu\~na Mackenna 4860, 782-0436 Macul, Santiago, Chile\\
\email{mcatelan@astro.puc.cl}
}

\authorrunning{Catelan}

\titlerunning{Multimodal Horizontal Branches}

\abstract{
We review the available empirical evidence for the presence   
of ``gaps'' and multimodal distributions among horizontal branch (HB) 
stars, along with some of the theoretical scenarios that have been 
proposed to explain these features. While gaps along the HB 
have become increasingly less prominent and frequent as more and 
better color-magnitude diagram data have been obtained for Galactic 
globular clusters, the evidence for multimodal HBs has instead become 
stronger. In addition, different HB modes have recently started to be 
traced down to multiple components that have been detected among 
subgiant branch and main sequence stars, thus suggesting that their 
origin lies in the complex physical processes that took place at the 
earliest stages in the history of massive stellar clusters. 

\keywords{Stars: Population II~-- stars: fundamental parameters~-- Hertzsprung-Russell 
  diagram~-- Galaxy: globular clusters~-- Galaxy: stellar content}
}
\maketitle{}

\section{Introduction}

``Gaps'' and ``multimodality'' are two terms that one will invariably come across,  
when studying the literature that addresses the horizontal branch (HB) morphology 
of resolved globular star clusters (GCs). In this sense, it is instructive to review 
the following authoritative definitions of the key terms ``gap'' and ``mode,'' as 
provided by the Cambridge dictionary: 

\begin{center} {\bf Gap:} {\em an empty space or opening in the middle of something 
or between two things}
\smallskip
 

{\bf Mode:} {\em the number or value which appears most frequently in a particular set} 
\end{center}

According to these definitions, a multimodal distribution may clearly exist 
{\em without} any gaps being present~-- all that is required, in this case, is 
for the probability distribution to present two or more statistically significant 
peaks. Indeed, as we shall soon see, the empirical 
evidence has increasingly been suggesting that ``empty spaces'' or ``forbidden regions'' 
do not actually exist along the HB. The evidence for HB 
multimodality, on the other hand, {\em is} becoming increasingly stronger, especially 
among some of the most massive globulars, driving renewed interest in theoretical 
interpretations of the observed features. 

It is interesting to note that the first 
reported gaps on observed color-magnitude diagrams (CMDs) were not located on 
the HB, but rather either along the main sequence (MS) or the red 
giant branch (RGB) of both open \citep{mj57,es64,es69} and globular 
\citep*{asea68,hr74} clusters. 

Not all such gaps have withstood the test of time. In particular, RGB 
gaps, once viewed as a ``major significant feature'' \citep{asea68}, were soon 
attacked on a statistical basis \citep{by72}. Later, 
\citet{rfp88} pointed out that ``the tendency has been for such gaps to get 
filled with increasing sample size''; it is indeed  
unusual for one to find recent CMD studies in which significant RGB 
gaps are claimed to be present. On the other hand, some of the 
gaps along the MSs of open clusters 
\citep[the ``B\"ohm-Vitense gaps'';][]{ebv70,bvc74} are still widely thought 
to be real \citep[e.g.,][]{rc00}, and caused by the change in behavior of 
convection as a function of MS mass \citep{fdea02b}. 
Will any of the widely reported HB gaps similarly withstand the test of time, 
or will they share the same fate as the RGB gaps?

\section{Gaps along the HB}

Apparently the first HB gap to have been identified 
was that along the blue HB of M12~= NGC\,6218 \citep{rr71}. 
\citeauthor{rr71}, in addition, remarks that similar features may be present in 
several other globulars, including NGC\,4147, M2 (NGC\,7089), M13 (NGC\,6205), M15
(NGC\,7078), M22 (NGC\,6656), and M92 (NGC\,6341). In fact, it is quite curious  
that \citet{asea68}, while calling attention to what they termed ``a major 
significant feature'' (i.e., a gap) along the RGB of M15, should have missed 
the gap on the blue HB of the cluster, which was to be prominently emphasized  
(and thereafter to play quite an influential role in shaping ideas in this 
field) almost two decades later \citep*{rbea85}. HB gaps have become 
mainstream mainly after the work by \citet{bn73} and \citet{ng76}, who 
identified two gaps 
along the color-color diagram of {\em field} blue HB stars, located at 
$T_{\rm eff} \simeq 12,\!900$~K and at $T_{\rm eff} \simeq 21,\!900$~K~-- the 
so-called {\em Newell gaps} 1 and 2, respectively~-- and by \citet{ns78}, who 
identified a similar gap to Newell's gap 2 along the extended blue HB of NGC\,6752, 
based on photometry by Cannon \& Lee (1973, unpublished; see Fig.~3 in 
\citeauthor{lc80} \citeyear{lc80} for their original NGC\,6752 CMD). 
Subsequently, many other 
clusters have been claimed to show signs of gaps along the HB 
\citep[see][ for a review and extensive references]{mcea98}. 
Are any such gaps real, and, if so, which? 

\begin{figure*}[t!]
\center
\resizebox{8cm}{!}{\includegraphics[clip=true]{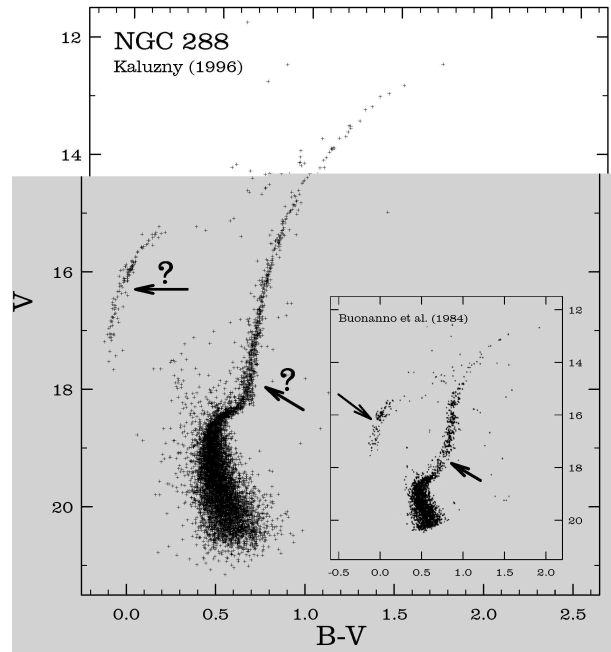}}
\caption{\footnotesize
Comparison of the recent NGC\,288 CMD by \citet{jk96} with the original one by 
\citet[][ {\em inset}]{rbea84}. Arrows indicate gaps in the 1984 CMD that appear 
to have vanished in the more recent of the two studies. 
}
\label{fig:N288}
\end{figure*}

It is 
very difficult to provide a conclusive answer to this question. At least some
of the reported gaps will likely vanish, or at least 
become much less prominent, with increasing sample sizes, as in the case of the RGB 
gaps that historically preceded them. Such a tendency was already noted a decade ago 
by \citet{mcea98}, who called attention to the fact that recent photometry has tended 
to cast doubt on the reality of at least some of the gaps which were originally 
reported to be present along the CMDs of GCs. As an example, we 
show, in Figure~\ref{fig:N288}, a comparison between the NGC\,288 CMD obtained by 
\citet{rbea84} and the one obtained by \citet{jk96}. The hotter of the two Newell 
gaps identified in NGC\,6752 by \citet{ns78}, which appeared as a 1~mag-wide 
void in the original CMD by \citet{lc80}, has with time also 
proved not to be devoid of stars as originally thought, but rather a region of 
the CMD that simply appears to be more sparsely populated than its surroundings 
\citep{rbea86,itea99}: bona-fide, spectroscopically confirmed HB stars are indeed  
present in its interior \citep*{smea97}. 

Several authors, on the other hand, have suggested that HB gaps are not only 
ubiquitous in the CMDs of GCs, but in fact are also located at the same place 
(as defined by either $T_{\rm eff}$ or total mass) in all clusters 
\citep[e.g.,][]{ffea98,vc99,gpea99,ymea04}. 
However, it should be noted that 
there are several published, high-quality CMDs for GCs with 
well-developed blue HBs that reveal {\em no} obvious gaps of any sort. 
Figure~\ref{fig:N288} shows NGC\,288 to be one such cluster~-- but other noteworthy 
cases include M79 \citep[NGC\,1904;][]{vdea96} and M2 \citep[NGC\,7089;][]{lc99}. 
We conclude that much more extensive, high-precision, multi-band, wide-field 
photometric surveys of GCs will be needed before we are in a 
position to conclusively establish how commonplace HB gaps really are. 

In addition to more extensive photometry, more robust statistical techniques 
are also required to test the reality of features (such as gaps) in the observed 
CMDs: as shown by \citet{mcea98}, the widely employed recipes provided by 
\citet{th71} give misleading results, generally overestimating the statistical 
significance of the detected features.\footnote{As an example, \citet*{dcea88} 
estimated, on the basis of the \citet{rbea84} CMD, a 100\% probability that the 
gap on the blue HB of NGC\,288 was {\em not} a statistical fluctuation~-- which 
is clearly inconsistent with the evidence presented in Figure~\ref{fig:N288}.}
Possibly a combination of the \citet{by72} 
approach with the binomial formalism proposed by \citeauthor{mcea98} could 
lead to more realistic results.

\section{Whence the ``Newell Gaps''?}\label{sec:newell} 

\cite{bb03b} has recently carried out an extensive 
study of rotation velocities of field HB stars, and included 27 stars from the 
\citet{bn73} and \citet{ng76} samples. On the basis of high-resolution 
spectra obtained with the McDonald Observatory's 2.1m telescope, \citeauthor{bb03b} 
concluded that {\em ``fewer than half (11 of 27) of the Newell stars that we observed 
were clearly HB objects, with another 11 stars classified as Population~I dwarfs, 
and the remaining five stars marked as pAGB [post-Asymptotic Giant Branch], 
subgiants, and such.''} With such a contaminated sample, one cannot help but 
wonder ``why would gaps appear in the color distribution of such 
a heterogeneous set of stars.'' In view of such evidence, another careful 
look at the data for field blue HB stars is clearly in order.

\section{HB Gaps and HB Theory} 

Canonical HB theory does not predict any sharp transitions along the HB. 
Accordingly, gaps are not naturally predicted by theory, unless one assumes 
sharp discontinuities in envelope mass~-- or, equivalently, in the amount of 
mass loss among red giants \citep[e.g.,][]{rtr98}. 

The suggested invariance in the position of certain HB gaps from one cluster 
to the next \citep{ffea98,gpea99}, irrespective of [Fe/H] or central density, 
would require a fine tuning in the RGB mass loss process, since mass loss is 
expected to depend on both metallicity and stellar interactions. Therefore, 
for any such gaps to be consistently present in a range of globular clusters 
with different [Fe/H] or/and central densities, one would most likely need 
to invoke scenarios in which mass loss on the RGB is {\em not} the 
culprit, but rather atmospheric phenomena operating, for instance, at the 
transition between radiatively levitated and ``normal'' blue HB stars 
\citep[e.g.,][]{vc99,fgea99}. 

Indeed, even for a mass distribution along the zero-age HB (ZAHB) containing 
true gaps, the corresponding canonical CMD will most likely end up containing 
no gaps at all, as a consequence of evolution away from the ZAHB. In this sense, 
\citet{vc99} has pointed out that a real gap should not be seen in the CMDs of 
GCs at $\approx 10,\!000$~K, {\em unless} the HB mass distribution 
presented a gap as large as $0.07\,M_{\odot}$. This is indeed a large 
gap in mass: according to the models by \citet{mcea98} for a metallicity 
$Z = 5\times 10^{-4}$, it would be sufficient to move a ZAHB star from 
around the middle of the RR Lyrae instability strip 
to a position that is some 4000~K hotter, well into the blue HB. 
Clearly, it is not a trivial matter to produce a true gap along the HB, 
at least for typical GC metallicities. The scenario whereby increasingly 
hotter HB gaps are produced as a consequence of successive mass loss episodes 
on the RGB, possibly related to stellar encounters 
(or even to planet engulfment; \citeauthor{ns98} \citeyear{ns98}), also 
suffers from the fact that the relative proportions of stars in between each 
successive (hotter) gap is not consistent with the expectations; from the 
lack of noteworthy differences in the radial distribution of HB stars with 
different colors; and from the requirement that the encounters must be fine 
tuned to produce mass loss amounts that are nearly the same from one star 
to the next \citep{csea97}. Finally, and as also noted by 
\citeauthor{csea97}, stellar encounters can certainly not be the cause 
of the Newell gaps among field HB stars (but see \S\ref{sec:newell}).

\subsection{HB Gaps and the Production of EHB Stars} 
The situation 
changes dramatically at very high metallicities though, when the range in mass 
between the red edge of the RR Lyrae instability strip and the extreme HB (EHB)
becomes narrower, with the net effect that a small gap in mass can indeed lead 
to a large gap {\em both} in broadband colors {\em and} in effective temperatures
\citep*[e.g.,][]{bdea93,ndcea96,syea97}. As shown by \citeauthor{ndcea96}, 
this result becomes especially clear when one carries out CMD simulations in which, 
instead of using RGB mass loss as a free parameter, one uses instead a {\em mass 
loss efficiency parameter} as the free parameter. In fact, such an approach also 
helps remove the ``fine-tuning problem'' for the production of EHB stars that has 
long puzzled astronomers.\footnote{Unfortunately, it remains unclear what mass loss 
recipe should 
be used for first-ascent RGB stars \citep[see][ for a critical discussion]{mc07b}. 
Very recently, \citet{loea07} carried out Spitzer observations of red giants in 
GCs, and concluded that mass loss is episodical and better 
described by a mass loss ``law'' with a very mild dependence on the evolutionary 
parameters of the stars (such as radius, luminosity, and gravity).} 
The fact that EHB stars, but not ordinary blue HB stars, can indeed be 
produced at very high metallicities is dramatically demonstrated by the CMD of the 
old, supersolar-$Z$ open cluster NGC\,6791, which possesses a well-developed 
red HB co-existing alongside several stars on the {\em EHB}~-- with basically 
nothing in between the two groups \citep[e.g.,][]{kr95,wlea98,gcea06}. 

In regard to the physical origin of the EHB stars~-- which are the field counterparts 
of the blue subdwarf (sdB) stars in the field~-- it is important to note that, while 
a large fraction of field sdB stars appears to be in (close) binary systems 
\citep*[e.g.,][]{pmea01,rnea04}, the same has recently been found {\em not} 
to be the case in GCs, where EHB stars in close binary systems 
seem to be rare \citep*{mbea06,mbea07,mbea08}. This strongly suggests 
that, while binary interactions may be involved in the production of a sizeable 
fraction of field EHB stars, single-star mechanisms may be more efficient in 
the case of GCs. In fact, recent results suggest that the MS 
binary fractions in the outskirts of the GCs NGC\,6397 and 
NGC\,6752 are very low, thus indicating, according to the results of realistic 
$N$-body simulations by \citet*{jhea07}, that the {\em primordial} binary fraction 
in GCs may also have been surprisingly low \citep{hrea07,mcea08}. 
If confirmed by more extensive observations of an enlarged sample of GCs, 
this may considerably limit the usage of EHB 
stars in globulars as indicators of what should be expected in integrated-light 
far-UV observations of distant field populations, including the UV upturn 
phenomenon in elliptical galaxies \citep{mc07b}.

\subsection{Plausible Physical Mechanisms}\label{sec:physical}
To be sure, certain physical mechanisms 
have been identified which may plausibly give rise to real HB gaps, 
at least in {\em some} observational planes. This 
is the case, in particular, of radiative levitation/gravitational diffusion, 
whose onset is quite abrupt at $T_{\rm eff} \simeq 11,\!500$~K, and which has 
been identified as the root cause of the so-called ``Grundahl jump'' that is 
observed in CMDs where either Str\"omgren's $u$ or Johnson's $U$ bands are 
used \citep{fgea99}. Indeed, all stars hotter than the jump are now known 
to have chemical compositions that are vastly different from the original 
ones, whereas the cooler stars have compositions that are representative 
of the star's original mix 
\citep*[e.g.,][ and references therein]{pbea95,fcea97,bb03a}. This abrupt 
discontinuity in chemical abundances can plausibly lead to observable CMD 
features, such as the Grundahl jump itself. Since the radiative 
levitation phenomenon occurs for {\em all} HB stars hotter than 
$\simeq 11,\!500$~K, one should accordingly expect that 
any related features in the CMD should be seen in all objects containing 
sufficiently hot HB stars as well. 

This may be the greatest difficulty facing the association between the 
radiative levitation phenomenon and the ``G1'' gap of \citet{ffea98} 
\citep[see also][]{vc99}. Indeed, while the Grundahl jump {\em is} 
indeed ubiquitous \citep{fgea99}, it remains unclear whether the G1 gap 
is present in {\em all} GCs with sufficiently hot HB stars. 
In fact, according to the calculations carried out by \citeauthor{fgea99}, 
one expects the large abundance increases brought about by radiative 
levitation to have a maximum impact upon CMDs in which $u$ or $U$ is used.  
On the other hand, in the visual and in the near and far UV, the effects 
are expected to become much 
smaller.\footnote{To be sure, the calculations carried out by 
\citet{fgea99} assumed an overall enhancement of {\em all} metals to supersolar 
levels, whereas the observations indicate a wide range of enhancement levels 
\citep[e.g.,][]{pbea95,fcea97}. To the best of our knowledge, no calculations 
have been carried out so far in which the observed element ratios were properly 
taken into account~-- and this would be of great interest in the present 
context. In like vein, 
it remains to be seen whether the low gravities that are 
commonly observed over the same $T_{\rm eff}$ range as the Grundahl jump can 
be entirely explained in this way, or whether additional physical mechanisms, 
such as He mixing on the RGB, may also be required \citep[e.g.,][]{smea03}.}

Note that the Grundahl jump is also accompanied by a 
discontinuity in the rotation velocities \citep[e.g.,][]{bb03a,arbea04}: 
blue HB stars hotter than 11,500~K present essentially no rotation, in 
contrast with cooler stars which may show significant rotation 
velocities, up to $\sim 40\,{\rm km}\,{\rm s}^{-1}$. 
\citet{as02} suggests that the low rotation velocities of stars hotter 
than the Grundahl jump may be due to the spin down of the surface layers 
by a weak stellar wind induced by the radiative levitation of Fe 
\citep[but see also][]{bs08}. 
As shown  by \citet{vc02}, such winds are indeed predicted by theory. 

At higher temperatures, another feature has recently been identified 
and claimed to be ubiquitous (among GCs containing sufficiently 
hot HB stars), namely the so-called ``Momany jump,'' 
at $T_{\rm eff} = 21,\!000 \pm 3000$~K \citep{ymea04}. As in the 
case of the Grundahl jump, this jump becomes more evident in  
CMDs that use the $u$ or $U$ bandpasses. 

While its physical origin is not yet clear, \citet{ymea04} suggested 
that this feature is somehow related to the {\em hot flashers}, 
which are stars that have lost so much mass during their 
RGB evolution that they fail to ignite He at the RGB proper, the He flash  
taking place after the star has ``peeled off'' the RGB 
\citep{cc93,ndcea96,tbea01,scea03}. However, the fact that the hot 
flashers appear at $\simeq 35,\!000$~K, which is much hotter than the Momany 
jump, casts some doubt on this interpretation. As in the 
case of the Grundahl jump, another explanation could be that 
abrupt chemical composition changes, again related to diffusion/levitation 
effects, take place at $\approx 21,\!000$~K, thus giving rise to the suggested 
feature. 

There are two types of hot 
flashers~-- {\em early} and {\em late} flashers. The former succeed 
in igniting He prior to the He white dwarf (WD) cooling curve, whereas the latter
ignite He only after the star has started climbing down this curve. Most 
interestingly, late flashers, as opposed to the early flashers, are predicted 
to undergo extensive mixing 
between the He core and the envelope, thus becoming enriched in He and C. 
As a consequence, early flashers are predicted to arrive at the ZAHB at a 
position near the end of the canonical EHB, whereas the ZAHB position for the 
late flashers becomes over 5000~K hotter (and somewhat fainter; see, e.g., 
Fig.~9 in \citeauthor{tbea01} \citeyear{tbea01}, or Fig.~2 in \citeauthor{smea02} 
\citeyear{smea02}), giving rise to the so-called ``blue hook'' that is observed 
in far-UV GC CMDs. Whether the predicted gap in ZAHB 
temperatures and surface chemical composition translates into a real gap 
being observed on the CMD again depends on the exact bandpasses used 
\citep[see, e.g., Fig.~16 in][]{tbea01}; also, a gap may be at least partly 
washed out by the effects of evolution away from the ZAHB. Unfortunately, 
there are not many GCs with sufficiently hot HB stars that the late 
flasher predictions can be verified on the basis of large  
samples: at present, the only GCs with confirmed blue hook stars 
are $\omega$~Cen \citep{jwea98,ndcea00}, M54~= NGC\,6715 \citep{area04}, NGC\,2419 
\citep{vrea07}, NGC\,2808 \citep{tbea01}, NGC\,6388, and (possibly) NGC\,6441 
\citep{gbea07}.  
However, the available evidence seems rather encouraging, and indeed appears 
to favor the late flasher scenario for the production of extremely hot HB 
stars beyond the classical EHB limit, over such alternatives as the 
He polution scenario \citep{smea02,smea07,tlea04}. Finally, RGB stars that 
lose even more mass than the blue hook progenitors will miss the HB phase 
completely, evolving directly down the WD cooling curve and never igniting
He in their cores, thus becoming He WDs \citep[e.g.,][]{jkea07,vcea07}.

\section{HB Multimodality} 
While the empirical evidence for real gaps may remain somewhat dubious,  
HB multimodality seems to rest on a much stronger footing. At least in a few 
clusters, the evidence for two or more modes on the HB has recently been traced 
down to multiple populations on the subgiant branch (SGB) or/and the MS. 

Even then, one should still be careful to avoid overinterpreting the data for HB 
stars. Indeed, depending on the type of observations carried out, modes can 
be naturally generated {\em without} any physical parameter of the HB stars 
presenting a multimodal 
distribution. This is well known to happen in the case of optical CMDs;
here, a continuous and uniform mass distribution can easily lead to a bimodal 
distribution in HB colors \citep[see Fig.~15 in][]{mcea98}, simply because of the 
saturation of optical colors for the hot blue HB stars. Quite often, to avoid 
confusion, multi-band photometry, including the near- and far-UV, is needed to 
better track the variation in the stellar physical parameters, such as mass,  
effective temperature, and gravity, along the HB \citep[e.g.,][]{ffea98,gbea07}. 

NGC\,2808 is by far the best documented case of a GC with a multimodal 
HB~-- but it is unlikely to be the only one. \citet{mcea98} classify as ``bimodal 
HBs'' all those GCs which present a deficit in the RR Lyrae number counts, 
compared to both red and blue HB stars. NGC\,2808 has long been known to have 
a well-populated red HB component coexisting with a blue HB, with little in 
between~-- i.e., at the RR Lyrae ``gap'' \citep[]{wh74}.\footnote{As pointed 
out by \citet{mc05}, the term RR Lyrae ``gap'' is very inadequate, but is 
still commonly used. This is because, in order to properly place 
an RR Lyrae in a CMD, one needs to follow its 
whole pulsation cycle and thereby obtain reliable mean colors and magnitudes. 
Since most CMD studies lack adequate time coverage, 
these variable stars are often simply omitted from the published CMDs, thereby 
leading to an {\em entirely artificial} empty space, or ``gap,'' between the 
red and blue HB components.} 
More recent photometry 
has revealed that NGC\,2808 actually {\em does} contain a significant RR Lyrae 
component, though with many fewer stars than either its red or 
blue HB counterparts \citep{mcea04}. 
In addition, deeper wide-field studies, as well as 
high-resolution HST photometry of the innermost cluster regions, have revealed 
an amazing superposition of what appear to be well-defined modes 
along the blue HB of the cluster \citep[e.g.,][]{lbea00,vcea06}. 

Very recently, it has been shown that the NGC\,2808 MS is actually comprised 
of three distinct components, which are more straightforwardly explained as three 
different populations with different helium abundances but nearly the same 
metallicities \citep{fdea05,gpea07}. It is, of course, very tempting to 
associate these different MSs to the different HB components that 
are present in the cluster, and \citeauthor{gpea07} point out that the different 
proportions of stars along the main branches of the cluster appear tantalizingly  
consistent with this notion. 

The association of abundance anomalies 
with HB morphology was originally advanced by \citet{jn81},
\citet{jnea81}, and \citet{sn83} 
in the context of CN variations, and by \citet{cfp95} in the context 
of super oxygen-poor stars. More recent studies include, among others, those by 
\citet{ecea07} and \citet{dv07}. The observation of abundance anomalies among 
{\em unevolved} stars in GCs \citep[e.g.,][]{rgea01} has 
given strong support to the notion that at least some of the variations more 
frequently observed among RGB stars dates back from the earliest stages in the 
lifes of these clusters, although deep mixing effects may still play a relevant 
role in explaining some of the abundance patterns observed in giants 
\citep[e.g.,][]{csea04,dv07}. Still, it remains unclear how high 
levels of He enrichment can be produced among GC stars without an 
accompanying change in metallicity
\citep[see, e.g.,][]{akea06,kbea07a,kbea07b,cy07}. 
In any case, it must be noted that the origin of the hottest stars lying on the 
extension of the EHB~-- the blue hook stars (\S\ref{sec:physical})~-- {\em cannot} 
be entirely explained in terms of the high He scenario, their observed properties 
being instead most consistent with the late-flasher scenario \citep{smea07}. 

It remains to be seen how many GCs will present convincing 
evidence for primordial abundance variations, since most globulars 
still appear to be well described within the framework of simple stellar 
populations (Piotto 2008, this volume). 
Indeed, only the most massive globulars have been found 
or suggested to contain composite populations; so far the evidence for 
multiple MSs or/and SGBs remains restricted to the 
cases of $\omega$~Cen, NGC\,2808 \citep{gpea07}, and NGC\,1851 
\citep{amea07}~-- though other massive clusters, such as NGC\,6388 and NGC\,6441, 
are also suspected of harboring heterogeneous populations, with a direct 
impact upon their observed HB morphologies \citep[e.g.,][]{gbea07,cd07}. 

It is important to note that 
the three GCs for which composite CMDs have been conclusively established 
(i.e., NGC\,1851, NGC\,2808, and $\omega$~Cen) all differ in important 
respects. More specifically, $\omega$~Cen appears to be affected by {\em both} 
metallicity {\em and} He abundance variations, with a large age spread also 
being present, whereas in NGC\,2808 no spread in age or metallicity 
has been detected. The SGB split 
observed in NGC\,1851 is formally consistent with two populations differing 
in age by $\sim 1$~Gyr; however, according to the models by 
\citet{mcea01} and \citet{mc05} in their study of the pair 
NGC\,288/NGC\,362, such an age spread would be insufficient to explain the 
HB bimodality observed in the cluster. Note, in addition, that the deep 
photometry by \citet{amea07} reveals a very tight MS, indicating that there 
is unlikely to be a sizeable metallicity or He spread in this GC. On 
the other hand, \citet{fg03} has noted that NGC\,1851 differs from NGC\,288 
and NGC\,362 in that it presents scatter in both the Str\"omgren $c_1$ and 
$m_1$ indices, and suggests that the $m_1$ spread in particular could be 
due to a spread in CN. It remains to be seen 
whether these CN variations would be able to explain the SGB split 
and bimodal HB of the cluster. 

To close, we note that another HB mode may be present at the very {\em red} 
end of the HB, comprised of {\em blue straggler star} (BSS) {\em progeny}. 
Indeed, one should naturally expect that BSS, once they evolve away 
from the MS and start burning He in their core, will become red clump 
stars, thus tending to be brighter, redder and more luminous than regular 
red HB stars (see \S\S2.1 and 2.2 in \citeauthor{mc05} \citeyear{mc05} for 
a recent discussion). These stars have been called ``evolved BSS,'' or 
E-BSS, by \citet{ffea99}. 
Therefore, in GCs containing large numbers of BSS, 
a small E-BSS component is expected to be present, with of order 
1 E-BSS star for every six or so BSS \citep{fpea92}. Indeed, 
such a component has been tentatively identified in several clusters 
\citep[e.g.,][]{fpea92,ffea99,sb04}.

\section{Conclusions}
The empirical evidence for HB gaps has become weaker over the past several 
years, with several previously reported such features having become at least 
partly filled as more and better data have become available. It does not 
appear very likely at present that {\em all} GCs have HB gaps. Still, 
plausible mechanisms for the production of some HB gaps have been advanced, 
generally involving a sharp discontinuity in surface chemical abundances with 
temperature. Such a discontinuity may be brought about by radiative levitation, 
which leads to the ``Grundahl jump'' at $T_{\rm eff} \simeq 11,\!500$~K, or by 
a late hot flasher, which leads to a predicted gap in ZAHB temperatures beyond 
the end of the canonical EHB. Such phenomena may lead to significantly different 
CMD features depending on the bandpasses used: in some planes, ``jumps'' and 
``gaps'' may become much more apparent than in others. 

In contrast, the empirical evidence for multimodal HBs has become significantly 
stronger lately. As a matter of fact, in several cases HB multimodality has 
been successfully traced to multiple sequences on the SGB or/and the MS, and 
in some such cases He enrichment, not necessarily accompanied by an increase 
in metallicity, is strongly suspected to be the culprit. HB multimodality 
does not appear to have a single origin though, and more data and theoretical 
studies are required before we are in a position to claim that we fully 
understand the origin of multimodal HB distributions.

\begin{acknowledgements}
I am indebted to J. Kaluzny for providing the NGC\,288 data in machine-readable 
format, and to A. V. Sweigart and H. A. Smith for critical readings of this paper 
and for useful comments. This research is supported by Proyecto Fondecyt Regular 
\#1071002.  
\end{acknowledgements}

\bibliographystyle{aa}

\end{document}